\documentclass{svproc}
\usepackage{url}

\usepackage{units}
\usepackage{graphicx}
\usepackage[hidelinks]{hyperref}
\usepackage{float}
\usepackage{amsmath}
\usepackage{amssymb}
\usepackage{subfigure}

\begin{document}
\mainmatter              
\title{Optimizing pulsed blowing parameters for active separation control in a one-sided diffuser using reinforcement learning}
\titlerunning{AFC Optimization with DRL}  
%
\author{Alexandra Müller$^1$ \and Tobias Schesny$^2$ \and Ben Steinfurth$^3$ \and Julien Weiss$^4$}
\authorrunning{Alexandra Müller et al.} 
%
\tocauthor{Alexandra Müller, Tobias Schesny, Ben Steinfurth, Julien Weiss}
\institute{Chair of Aerodynamics, Institute of Aeronautics and Astronautics,\\ Technische Universität Berlin, Marchstr. 12-14, 10587 Berlin}

\maketitle              

\vspace{-0.35cm}
\centerline{$^1$ \email{alexandra.mueller@campus.tu-berlin.de}}
\centerline{$^2$ \email{schesny@campus.tu-berlin.de}}
\centerline{$^3$ \email{ben.steinfurth@tu-berlin.de}}
\centerline{$^4$ \email{julien.weiss@tu-berlin.de}}

\begin{abstract}
Reinforcement learning is employed to optimize the periodic forcing signal of a pulsed blowing system that controls flow separation in a fully-turbulent $Re_\theta = 1000$ diffuser flow. Based on the state of the wind tunnel experiment that is determined with wall shear-stress measurements, Proximal Policy Optimization is used to iteratively adjust the forcing signal. Out of the reward functions investigated in this study, the incremental reduction of flow reversal per action is shown to be the most sample efficient. Less than 100 episodes are required to find the parameter combination that ensures the highest control authority for a fixed mass flow consumption. Fully consistent with recent studies, the algorithm suggests that the mass flow is used most efficiently when the actuation signal is characterized by a low duty cycle where the pulse duration is small compared to the pulsation period. The results presented in this paper promote the application of reinforcement learning for optimization tasks based on turbulent, experimental data. 


\end{abstract}
\section{Introduction}

Flow separation is often accompanied by an increase in drag and other unwanted behavior like stall on an airplane or surge in a compressor. This study focuses on active separation control, where an external energy source is required to manipulate the flow.
Compared to classical boundary layer control by means of steady suction or blowing, a higher control authority is achieved with oscillatory blowing.
Greenblatt and Wygnanski \cite{greenblatt2000} provide an overview of different means of active flow control by periodic excitation.
To quantify the relative momentum input associated with blowing in flow control applications, the dimensionless momentum coefficient $C_\mu = (c_\mu, \langle c_\mu \rangle)$ is defined as the ratio of added momentum over free-stream momentum. The momentum coefficient can be divided into $c_\mu$ for the steady and $\langle c_\mu \rangle$ for the oscillatory momentum.
Another parameter to describe means of active flow control is the dimensionless frequency $F^+=\nicefrac{(f \cdot x)}{U_\infty}$ with the excitation frequency $f$, a reference length $x$ and the free-stream velocity $U_\infty$.
In their experiment, Seifert et al. \cite{seifert1996} show that by periodically blowing over the flap of an airfoil with $F^+=2$, a smaller momentum input is needed for similar effect on lift and drag compared to the steady-blowing approach. This demonstrates the efficiency of the periodic excitation as well as the importance of the time scales defining the forcing signal.
In the present study, the active flow control is performed by Pulsed Jet Actuators (PJAs), similar to those used by Steinfurth et al. \cite{steinfurth2022} and Löffler et al. \cite{loeffler2023}.
The periodic forcing of a PJA is defined by two timescales, the pulse duration $t_\mathrm{p}$ and the time delay between successive pulses $t_\mathrm{off}$.



Recent advances in Machine Learning and Artificial Neural Networks (ANN) present new methods of optimizing active flow control parameters.
As pointed out by Rabault et al. \cite{rabault2020}, deep reinforcement learning (DRL) is a promising strategy for flow control optimization. This method is based on teaching empirical strategies to an ANN through trial and error. DRL already achieved improvements in the fields of robotics \cite{pinto2017}, language processing \cite{bahdanau2017} and games \cite{mnih2013}. It was first applied to an active flow control problem by Rabault et al. \cite{rabault2019}. The article shows that in a two-dimensional simulation, an ANN is able to learn an active flow control strategy by varying the mass flow rate of two jets, one on each side of a cylinder, thereby stabilizing the wake and reducing drag. 
Additional studies to control the flow around a cylinder were conducted by Ren et al. \cite{ren2021} who investigated the same problem as Rabault et al. \cite{rabault2019} but in weakly turbulent conditions. Fan et al. \cite{fan2020} discovered control strategies for a turbulent cylinder flow with two fast-rotating smaller control cylinders using DRL in experimental and simulation environments whereas Tokarev et al. \cite{tokarev2020} studied the application of DRL to control the laminar flow over an oscillating cylinder. Other DRL applications include flow separation control over an airfoil by Shimomura et al. \cite{shimomura2020} or suppressing vortex-induced vibration by Zheng et al. \cite{zheng2021}.

The present paper aims to apply DRL to optimize the active flow control parameters in the one-sided diffuser that was previously studied by Steinfurth et al. \cite{steinfurth2022} and Löffler et al. \cite{loeffler2023}. As opposed to the majority of previous studies, we consider experimental data of a fully-turbulent flow with a Reynolds number of $Re_\theta = 1000$ based on a momentum thickness $\theta \approx \unit[0.8]{\mathrm{mm}}$.
The objective is to optimize the periodic forcing parameters $t_\mathrm{p}$ and $t_\mathrm{off}$ of a PJA in order to reduce the turbulent separation bubble on the diffuser. 
The state is returned with wall shear-stress measurements along the diffuser. Actions consist in incremental changes of $t_\mathrm{p}$ and $t_\mathrm{off}$, and the suppression of reverse-flow is rewarded. During each episode, state-action-reward triplets are gathered for fifteen combinations of $t_\mathrm{p}$ and $t_\mathrm{off}$.
The main focus point of this study is the influence of different reward functions. Furthermore, the importance of explorative training is demonstrated.
The experimental setup and DRL algorithm are explained in \autoref{sec:methodology}. The results will be presented in \autoref{sec:results}, followed by a conclusion in \autoref{sec:conclusion}.




\section{Methodology} \label{sec:methodology}

The following section provides a detailed overview of the experimental setup and DRL algorithm used for this study.

\subsection{Experimental Setup}



\begin{figure}
    \vspace{-0.5cm}
    \centering
    \includegraphics[width=0.9\linewidth]{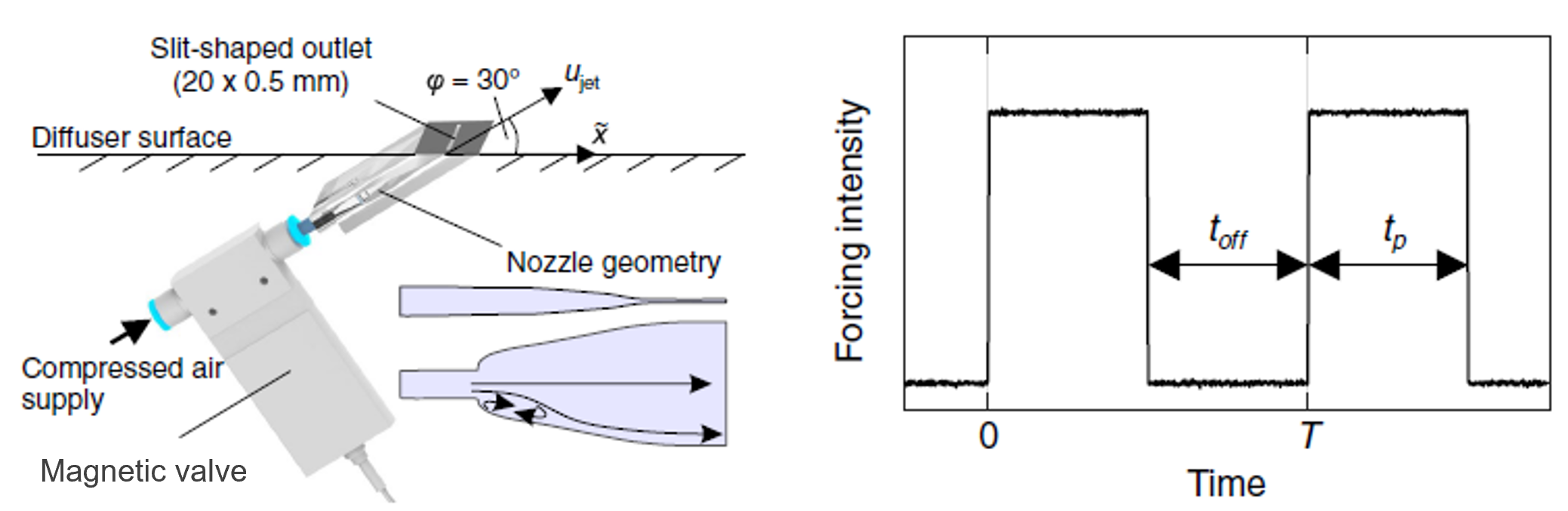}
    \caption{PJA for active separation control. Left: illustration of the actuator. Right: control signal \cite{steinfurth2022}}
    \label{fig:pja}
    \vspace{-0.5cm}
\end{figure}

\begin{figure}
    \centering
    \includegraphics[width=\linewidth]{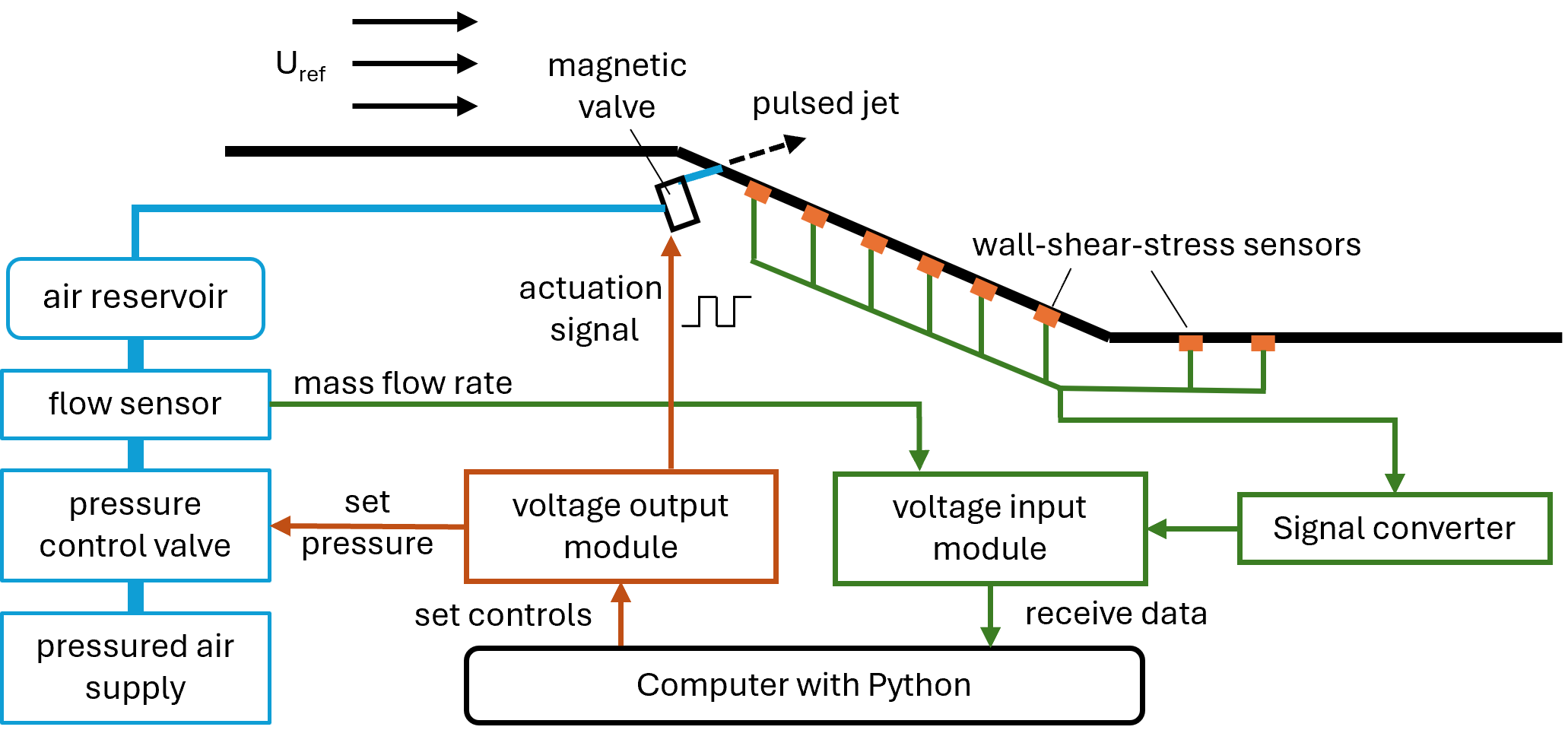}
    \vspace{-0.25cm}
    \caption{Schematic depiction of the experimental setup}
    \label{fig:exp_setup}
    \vspace{-0.5cm}
\end{figure}

The environment examined in this paper is a one-sided diffuser with a deflection angle of $\alpha=\unit[20]{^\circ}$ and a ramp length of $L=\unit[337]{\mathrm{mm}}$. The test section is $\unit[600]{\mathrm{mm}}$ wide and has a flat ceiling with a distance of $\unit[400]{\mathrm{mm}}$ to the bottom of the diffuser opening section. The diffuser is situated in a closed-loop wind tunnel with a nominal velocity of $U_{\mathrm{ref}}=\unit[20]{{\nicefrac{\mathrm{m}}{\mathrm{s}}}}$ and a Reynolds number of $Re_\theta = 1000$, based on a momentum thickness $\theta \approx \unit[0.8]{\mathrm{mm}}$. For active flow control, five PJAs with a $\unit[2]{\mathrm{mm}}$ spacing are embedded at the top of the diffuser ramp at an emission angle of $\varphi=\unit[30]{^\circ}$. Each PJA is connected to a magnetic valve that is controlled by a square wave signal. The square wave signal is composed of the pulse duration $t_\mathrm{p}$ and the time delay between successive pulses $t_\mathrm{off}$ (see \autoref{fig:pja}). The magnetic valves have a control uncertainty of $\pm \unit[15]{\mathrm{\%}}$ with regard to the pulse duration $t_\mathrm{p}$. The PJAs are connected to a compressed air supply with a fixed volume flow of $\dot{V} = \unit[60]{{\nicefrac{\mathrm{l}}{\mathrm{min}}}}$, which translates to a jet velocity $u_\mathrm{jet} = U_\mathrm{ref}$ when $t_\mathrm{p} = t_\mathrm{off}$. By finding the optimal actuation parameters for a fixed mass flow rate, the active flow control efficiency is maximized. The volume flow is measured by a flow sensor with a measuring range of $\dot{V} = 2$ - $\unit[200]{\mathrm{\nicefrac{\mathrm{l}}{\mathrm{min}}}}$ and an uncertainty of $\pm \unit[3.3]{\mathrm{\%}}$. The control parameter of the volume flow is the pressure provided by a pressure control valve. A schematic of the complete measurement setup is depicted in \autoref{fig:exp_setup}.

\begin{figure}
    \vspace{-0.5cm}
    \centering
    \includegraphics[width=\linewidth]{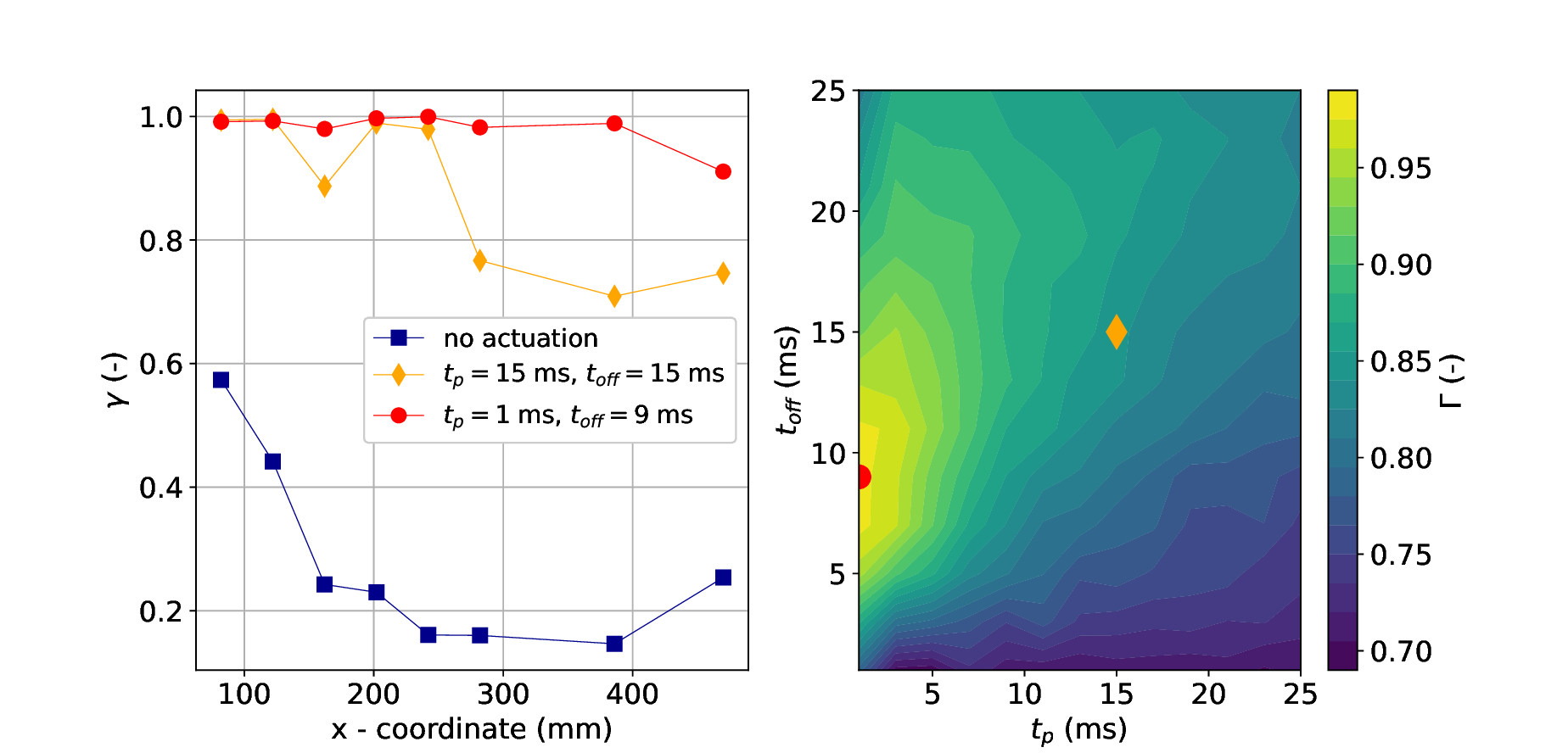}
    \vspace{-0.25cm}
    \caption{Measurement results. Left: distribution of the forward flow fraction $\gamma$ for different actuation parameters. Right: integrated forward flow fraction $\Gamma$ over entire parameter space}
    \label{fig:measurement_results}
    \vspace{-0.5cm}
\end{figure}

To measure the effectiveness of the active flow control, eight calorimetric wall-shear-stress sensors are embedded in the center line of the diffuser. With this setup, only the two-dimensional influence of the PJAs on the diffuser flow is observed. The used wall-shear-stress sensors were first introduced by Weiss et al. \cite{weiss2016,weiss2024}. They consist of an electrically heated microbeam suspended over a small cavity. Two additional lateral detector beams act as resistance thermometers by measuring the thermal wake of the heated beam. The sensors are sensitive to the flow direction and can be calibrated in a dedicated facility to further obtain the wall-shear-stress values. In this study, only the voltage output is required as it already provides information about the flow direction. At each sensor location, the forward flow fraction $\gamma$ is determined as the percentage of positive values of the signal relative to the total length of the signal. The signals are obtained at a frequency $f_\mathrm{s}=\unit[500]{\mathrm{Hz}}$ and a measurement time $T=\unit[10]{\mathrm{s}}$. 
As a scalar parameter for the PJA effectiveness, the normalized integral of the forward-flow fraction $\Gamma$ is defined as follows with $\Delta x$ being the distance between the first and the last sensor.
\begin{equation}
    \Gamma = \frac{1}{\Delta x} \int \gamma \; \mathrm{d} x
\end{equation}
To reduce the wall time required for training, but also to better understand the flow field, measurement data are gathered for a set of predefined values for $t_\mathrm{p}$ and $t_\mathrm{off}$. The DRL algorithm is then trained outside of the wind tunnel experiment by interpolating between the data points and thus making it a model-based approach. 
To construct this model of the system response, samples are taken in the range $t_\mathrm{p}, \; t_\mathrm{off} = 1$ - $\unit[25]{\mathrm{ms}}$ with an increment of $\Delta t=\unit[2]{\mathrm{ms}}$ as well as $t_\mathrm{p}=1$ - $\unit[7]{\mathrm{ms}}$ and $t_\mathrm{off}=3$ - $\unit[19]{\mathrm{ms}}$ with an increment of $\Delta t = \unit[1]{\mathrm{ms}}$. \autoref{fig:measurement_results} shows the integrated forward flow fraction $\Gamma$ over the entire parameter space. The parameter space contains one global optimum at $t_\mathrm{p}=\unit[1]{\mathrm{ms}}$ and $t_\mathrm{off}=\unit[9]{\mathrm{ms}}$ with a maximum integrated forward flow fraction $\Gamma = 0.98$. 
This is consistent with results by Steinfurth et al. \cite{steinfurth2022} and Löffler et al. \cite{loeffler2023}, where small $t_\mathrm{p}$ also proved to be effective.
Furthermore, \autoref{fig:measurement_results} depicts the distribution of the forward flow fraction $\gamma$ of a flow field with and without actuation. The latter suggest reverse-flow at the bottom of the ramp ($x = 250$ - $\unit[400]{\mathrm{mm}}$), which indicates the presence of a turbulent separation bubble. Actuation increases the forward flow fraction $\gamma$ in the whole flow field until almost all measurement positions indicate a $\gamma$ close to $1$, meaning a forward-facing flow in the wall-region, for the optimal actuation case. 

\subsection{Reinforcement Learning}

\begin{figure}
    \centering
    \includegraphics[width=0.6\linewidth]{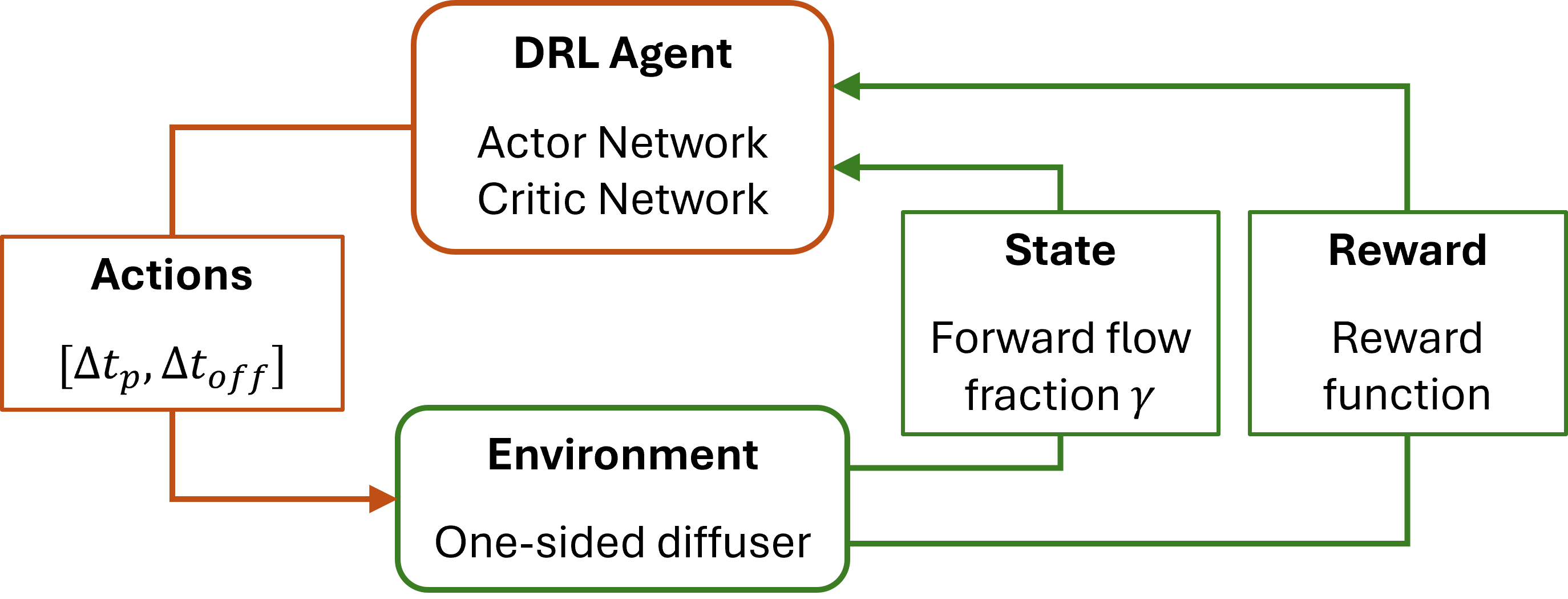}
    \caption{Schematic of one time step in Deep Reinforcement Learning}
    \label{fig:deep_reinforcement_learning}
    \vspace{-0.5cm}
\end{figure}

DRL is a machine learning methodology based on teaching empirical strategies to an ANN through trial and error.
It can be simplified as a cyclic process. The DRL agent, represented by one or more ANNs, performs an action based on a specific policy. The policy represents a probability distribution $\pi(a_t | s_t)$ that maps a state of the environment $s_t$ to an action $a_t$. 
The continuous action space spans incremental changes of the actuation parameters $a_t = [\Delta t_\mathrm{p}, \; \Delta t_\mathrm{off}]$. The actuation parameters at a time step $t$ are defined as:
\begin{equation}
    t_\mathrm{p}^t = t_\mathrm{p}^{t-1} + \Delta t_\mathrm{p}^t \qquad \text{and} \qquad
    t_\mathrm{off}^t = t_\mathrm{off}^{t-1} + \Delta t_\mathrm{off}^t
\end{equation}
The performed action has an effect on an environment, in this case the previously described one-sided diffuser. Observation of the environment provides a new state and reward to the DRL agent. The state of the environment is represented by an eight-element vector that includes the forward-flow fractions $\gamma$ along the diffuser test section. The reward function will be described later in this section. Based on the information about the state, the DRL agent generates a new action and the process is repeated. One loop of the process is called a time step and a sequence of time steps is called an episode. The DRL policy is updated at the end of each episode. A schematic of the process is shown in \autoref{fig:deep_reinforcement_learning}.

There are several algorithms that may be used as the DRL policy update. Viquerat et al. \cite{viquerat2022} provide a brief overview over the most commonly used DRL policies in the context of fluid mechanics. 
For this study, it was decided to use the Proximal Policy Optimization (PPO) due to its sample efficiency, robustness and easy application. PPO was first introduced by Schulman et al. \cite{schulman2017} and belongs to the Policy Gradient methods. Policy Gradient methods compute an estimator of the policy gradient and pass it to a stochastic gradient ascent algorithm \cite{schulman2017}. The approach in the present paper is model-based since the environment is modeled from experimental wind tunnel data. The algorithm is implemented in Python using the machine learning package PyTorch. The code is based on PPO implementations by \cite{henanmemeda2018} and \cite{tabor2020}.


The PPO algorithm requires two networks. The actor network provides an action based on an observation. It consists of an input layer with eight neurons, representing the state of the environment (i.e., the forward-flow fraction distribution), two hidden layers with 256 neurons, followed by an output layer with two neurons corresponding to the actions defined by $\Delta t_\mathrm{p}$ and $\Delta t_\mathrm{off}$. To ensure a continuous action space the actor network returns a Gaussian Normal distribution with a mean value $\mu$ between $\pm \unit[1]{\mathrm{ms}}$ and a pre-defined standard deviation $std$ for each element in the action space. The final action is then derived as a random sample from this distribution. The standard deviation can therefore be used as a dial to enforce exploration and exploitation as shown in \autoref{sec:results}.
The critic network learns to predict the value function which is used to evaluate the action chosen by the actor network. It has the same structure as the actor network, except for the output layer, which has only one neuron for the critic value.
Both networks are updated with a learning rate $lr = 3 \cdot 10^{-4}$.

The main focus of this study is to investigate the effect of three different reward functions $r$ on the optimization performance. 
The function $r = \Gamma$ returns the integral forward flow fraction $\Gamma$ of the current time step, whereas $r = \Delta\Gamma$ returns the difference in $\Gamma$ compared to the previous time step. The third reward is defined for each time step $t$ as the sum of integrated forward flow fraction $\Gamma$ of all previous time steps divided by the number of the current time step $t$:
\begin{equation}
    r = \Gamma_\mathrm{sum} = \frac{\sum_{i=0}^{t} \Gamma_i}{t}
\end{equation}
The three reward functions are investigated for different standard deviations of the action space $std = [\unit[0.4]{\mathrm{ms}},\; \unit[0.2]{\mathrm{ms}}, \; \unit[0.001]{\mathrm{ms}}]$. The standard deviation remains constant throughout the learning process. Each episode starts with the same initial actuation parameters $t_\mathrm{p} = t_\mathrm{off} = \unit[15]{\mathrm{ms}}$. The learning processes are run for 1,500 episodes with 15 time steps per episode.




\section{Results} \label{sec:results}

\begin{figure}
    \centering
    \includegraphics[width=\linewidth]{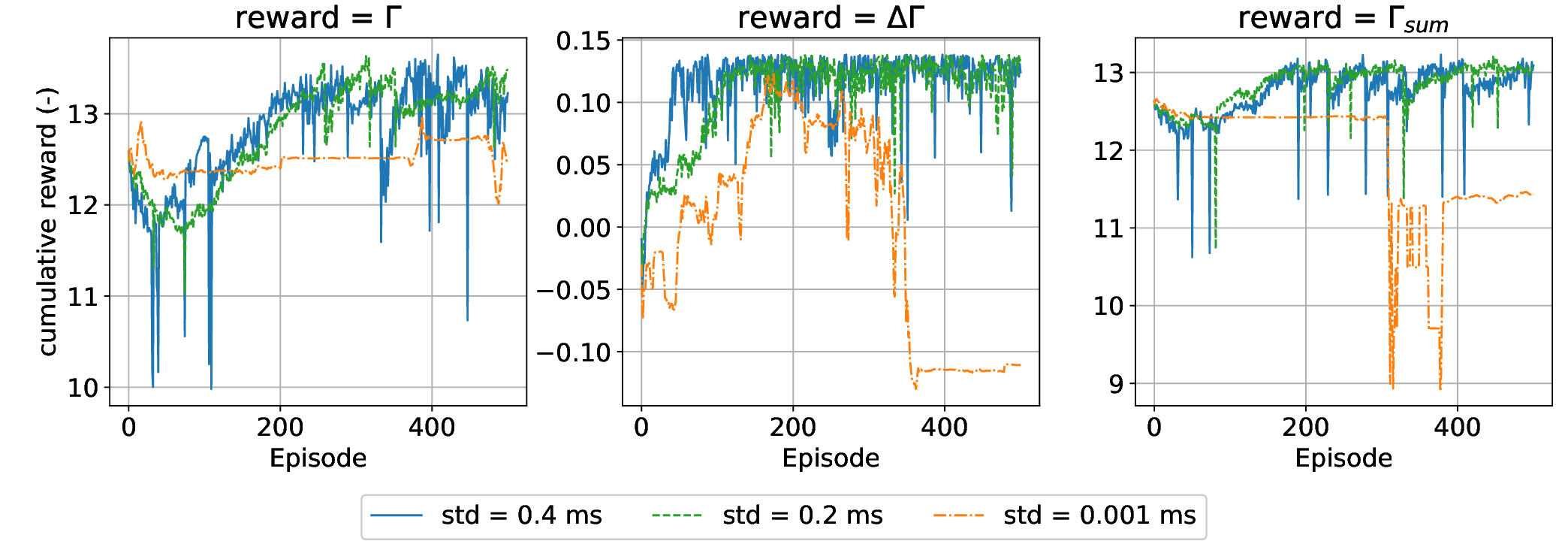}
    \vspace{-0.5cm}
    \caption{Cumulative reward per episode for different reward functions}
    \label{fig:cumulative_reward}
    \vspace{-0.65cm}
\end{figure}

\autoref{fig:cumulative_reward} shows the cumulative reward over the first 500 episodes for all reward functions and standard deviations.
The PPO algorithm aims to maximize the cumulative reward, which is defined as the sum of all rewards per time step in one episode. Thus, low rewards in single time steps are permitted, as long as the cumulative reward of the episode is high.
As explained previously, the standard deviation governs the width of the Normal Distribution from which the actions are derived, thereby enforcing either exploration for high standard deviations or exploitation for low standard deviations.
In \autoref{fig:cumulative_reward}, the blue and green curves depicting the standard deviations $std=\unit[0.4]{\mathrm{ms}}$ and $std=\unit[0.2]{\mathrm{ms}}$ increase for all three reward functions.
This indicates that the PPO agent develops an effective strategy for the flow control optimization.
The signal for the higher standard deviation $std=\unit[0.4]{\mathrm{ms}}$ is noisier and increases slightly faster for $r=\Gamma$ and $r=\Delta \Gamma$.
The orange curves depicting $std=\unit[0.001]{\mathrm{ms}}$ are smoother, but differ from the curves for the other standard deviations in all cases.
The cumulative reward increases and decreases for $r=\Delta \Gamma$ while not increasing at all for $r=\Gamma$ and $r=\Gamma_\mathrm{sum}$. It is also generally lower compared to the other standard deviations.
The low standard deviation $std=\unit[0.001]{\mathrm{ms}}$ equates to exploitation, meaning that the agent will always choose the "best" action without any random excursions.
An untrained algorithm first needs a lot of information about the parameter space to learn an effective strategy. Without exploration the chances that the algorithm "accidentally" pursues the correct direction is much lower.

\begin{figure}
    \centering
    \includegraphics[width=\linewidth]{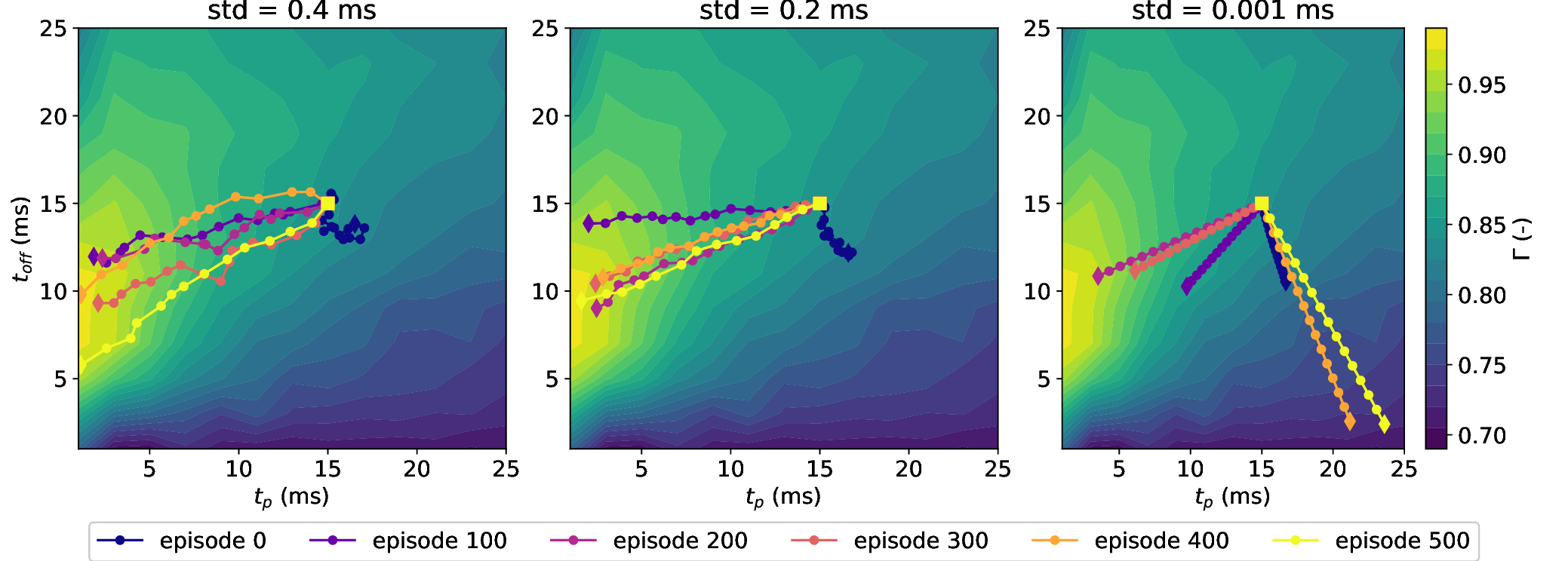}
    \vspace{-0.5cm}
    \caption{Progression of the action space $[t_\mathrm{p},\;t_\mathrm{off}]$ during specific episodes for different standard deviations $std$. The displayed reward function is $r = \Delta \Gamma$. The initial time step of each episode is depicted as $\Box$ and the final time step as $\diamondsuit$.}
    \label{fig:timestep_progression}
    \vspace{-0.5cm}
\end{figure}

\begin{figure}
    \vspace{-0.5cm}
    \centering
    \includegraphics[width=\linewidth]{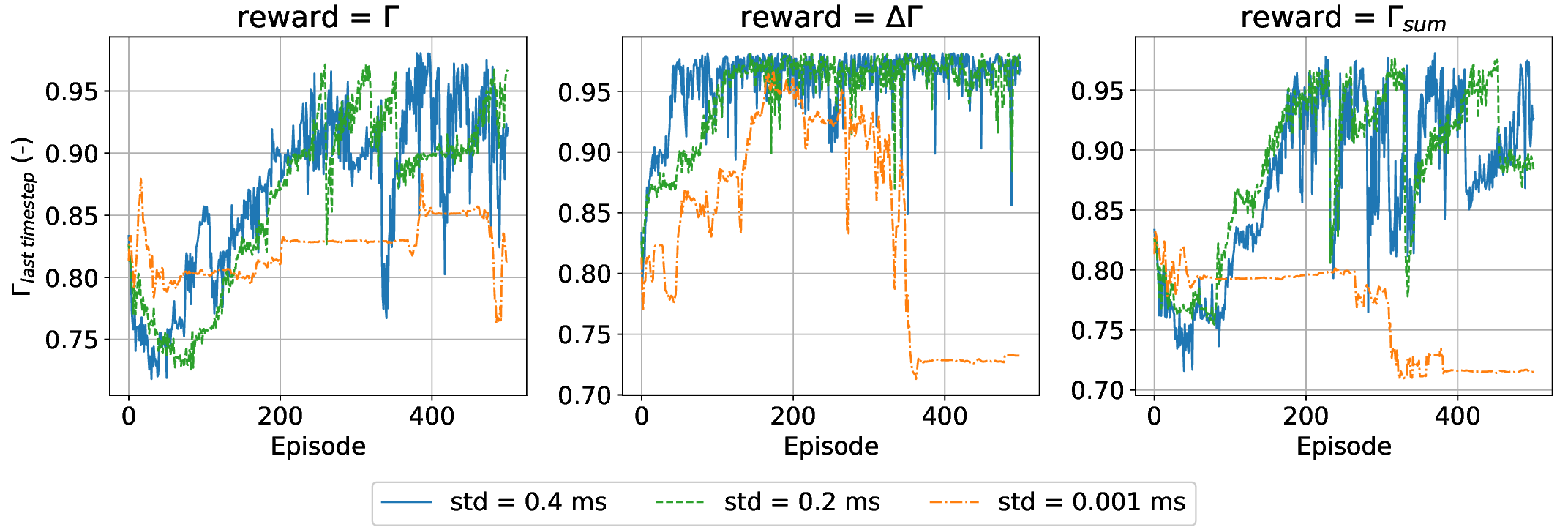}
    \vspace{-0.5cm}
    \caption{$\Gamma$ at the last time step of each episode for different reward functions}
    \label{fig:gamma_last_timestep}
    \vspace{-0.75cm}
\end{figure}

The influence of the standard deviation is also highlighted in \autoref{fig:timestep_progression} as an example for the $r = \Delta \Gamma$ function. It depicts the progression of the actuation parameters $t_\mathrm{p}$ and $t_\mathrm{off}$ during specific episodes throughout the learning process. For $std=\unit[0.4]{\mathrm{ms}}$ and $std=\unit[0.2]{\mathrm{ms}}$, the algorithm seems to pursue a random direction in the first episode, because of the noise added to the choice of the action with a non-negligible standard deviation. After a few hundred episodes, the algorithm repeatably finds the region of high forward flow fraction and therefore high reward. For $std=\unit[0.001]{\mathrm{ms}}$, the algorithm only follows a single direction in each episode. Although the actions are approaching the region of high reward in the episodes 200 and 300, the algorithm does not seem to learn an effective policy. A future approach, that is also pursued in other studies, could be to first run the training to a high standard deviation for a small number of episodes for fast learning and then switch to a lower $std$ for the rest of the learning process to reduce the noise and randomness. 
This study aims to optimize the flow control parameters by maximizing the forward flow fraction. To see if the PPO agents have reached this goal, \autoref{fig:gamma_last_timestep} depicts $\Gamma$ at the last time step of each episode. The distributions are very similar to that of the cumulative reward with $\Gamma$ increasing for $std=\unit[0.4]{\mathrm{ms}}$ and $std=\unit[0.2]{\mathrm{ms}}$.
Starting from an initial $\Gamma=0.84$, the algorithms for $r=\Gamma$ and $r=\Gamma_\mathrm{sum}$ reach values around $\Gamma = 0.9$ after 200 episodes. The algorithm for $r=\Delta \Gamma$ achieves values around $\Gamma=0.96$ after 50 episodes for $std=\unit[0.4]{\mathrm{ms}}$ and 150 episodes for $std=\unit[0.2]{\mathrm{ms}}$. Similar to the cumulative reward, the curves for the integral forward flow fraction decrease or stay constant for a standard deviation of $std=\unit[0.001]{\mathrm{ms}}$.

Comparing the three different reward functions reveals that the fastest learning with the highest achieved integral forward flow fraction $\Gamma$ can be accomplished with a reward function $r = \Delta \Gamma$. This can be explained as the parameter space only contains one optimum and the DRL algorithm quickly learns to follow the steepest gradient. 

\section{Conclusion} \label{sec:conclusion}
%


This paper demonstrates the optimization of active flow control parameters in a fully-turbulent one-sided diffuser flow based on experimental data using DRL with the PPO algorithm. The forcing parameters $t_\mathrm{p}$ and $t_\mathrm{off}$ of five PJAs embedded at the top of the diffuser ramp are adjusted while an array of wall-shear-stress sensors delivers the distribution of the forward flow fraction, that defines the state of the diffuser flow.
An investigation of the parameter space reveals an optimum in $\Gamma$ at a low pulse time $t_\mathrm{p}$, which is equivalent to a low duty cycle $DC = \nicefrac{t_\mathrm{p}}{(t_\mathrm{p} + t_\mathrm{off})}$. These results are consistent with the studies by Steinfurth et al. \cite{steinfurth2022} and Löffler et al. \cite{loeffler2023}. The former conducted a conventional parameter study and the latter used surrogate modeling to optimize the active flow control parameters.

The DRL is performed for three different reward functions at three different standard deviations. For higher standard deviations, all reward functions yield reasonable optimization strategies when starting the episode from fixed initial values for $t_\mathrm{p}$ and $t_\mathrm{off}$. Lacking the needed exploration at the beginning of the training, the algorithms trained with a small standard deviation do not perform well in comparison. 
Out of the three reward functions $r=\Delta \Gamma$ proves to be the most efficient.
This can be explained by the parameter space containing only one optimum.
In fact, other techniques may be viewed as more suitable for the task at hand. Nonetheless, applying DRL to this case does not only increase the knowledge on this specific topic, but also prepares for the next investigation approaches.

So far, each learning episode starts at the same initial $t_\mathrm{p}$ and $t_\mathrm{off}$ values. Thinking about possible active flow control applications in a real, non-testbed environment, a DRL algorithm might be subject to variable input parameters. The next step is to examine how well the algorithm performs when the actuation parameters are randomly initiated at the beginning of each episode. 

In summary, this study shows that training a DRL algorithm to perform flow control based on experimental data is possible while also pointing out topics for further investigation and improvement.



%
%
\bibliographystyle{IEEEtran} 
\bibliography{ref} 

\begin{thebibliography}{10}
\providecommand{\url}[1]{#1}
\csname url@samestyle\endcsname
\providecommand{\newblock}{\relax}
\providecommand{\bibinfo}[2]{#2}
\providecommand{\BIBentrySTDinterwordspacing}{\spaceskip=0pt\relax}
\providecommand{\BIBentryALTinterwordstretchfactor}{4}
\providecommand{\BIBentryALTinterwordspacing}{\spaceskip=\fontdimen2\font plus
\BIBentryALTinterwordstretchfactor\fontdimen3\font minus
  \fontdimen4\font\relax}
\providecommand{\BIBforeignlanguage}[2]{{%
\expandafter\ifx\csname l@#1\endcsname\relax
\typeout{** WARNING: IEEEtran.bst: No hyphenation pattern has been}%
\typeout{** loaded for the language `#1'. Using the pattern for}%
\typeout{** the default language instead.}%
\else
\language=\csname l@#1\endcsname
\fi
#2}}
\providecommand{\BIBdecl}{\relax}
\BIBdecl

\bibitem{greenblatt2000}
\BIBentryALTinterwordspacing
D.~Greenblatt and I.~J. Wygnanski, ``The control of flow separation by periodic
  excitation,'' \emph{Progress in Aerospace Sciences}, vol.~36, no.~7, pp.
  487--545, 2000. [Online]. Available:
  \url{https://www.sciencedirect.com/science/article/pii/S0376042100000087}
\BIBentrySTDinterwordspacing

\bibitem{seifert1996}
\BIBentryALTinterwordspacing
A.~Seifert, A.~Darabi, and I.~Wyganski, ``Delay of airfoil stall by periodic
  excitation,'' \emph{Journal of Aircraft}, vol.~33, no.~4, pp. 691--698, 1996.
  [Online]. Available: \url{https://doi.org/10.2514/3.47003}
\BIBentrySTDinterwordspacing

\bibitem{steinfurth2022}
\BIBentryALTinterwordspacing
B.~Steinfurth and J.~Weiss, ``Efficiency enhancement in active separation
  control through optimizing the duty cycle of pulsed jets,'' \emph{AIAA
  Journal}, vol.~60, no.~12, pp. 6566--6580, 2022. [Online]. Available:
  \url{https://doi.org/10.2514/1.J061667}
\BIBentrySTDinterwordspacing

\bibitem{loeffler2023}
\BIBentryALTinterwordspacing
S.~Löffler, B.~Steinfurth, and J.~Weiss, \emph{Surrogate-Based Exploration of
  Active Separation Control Parameters: An Experimental Study}, 2023. [Online].
  Available: \url{https://arc.aiaa.org/doi/abs/10.2514/6.2023-0076}
\BIBentrySTDinterwordspacing

\bibitem{rabault2020}
\BIBentryALTinterwordspacing
J.~Rabault, F.~Ren, W.~Zhang, H.~Tang, and H.~Xu, ``Deep reinforcement learning
  in fluid mechanics: A promising method for both active flow control and shape
  optimization,'' \emph{Journal of Hydrodynamics}, vol.~32, pp. 234--246, 2020.
  [Online]. Available: \url{https://doi.org/10.1007/s42241-020-0028-y}
\BIBentrySTDinterwordspacing

\bibitem{pinto2017}
\BIBentryALTinterwordspacing
L.~Pinto, M.~Andrychowicz, P.~Welinder, W.~Zaremba, and P.~Abbeel, ``Asymmetric
  actor critic for image-based robot learning,'' 2017. [Online]. Available:
  \url{https://arxiv.org/abs/1710.06542}
\BIBentrySTDinterwordspacing

\bibitem{bahdanau2017}
\BIBentryALTinterwordspacing
D.~Bahdanau, P.~Brakel, K.~Xu, A.~Goyal, R.~Lowe, J.~Pineau, A.~Courville, and
  Y.~Bengio, ``An actor-critic algorithm for sequence prediction,'' 2017.
  [Online]. Available: \url{https://arxiv.org/abs/1607.07086}
\BIBentrySTDinterwordspacing

\bibitem{mnih2013}
\BIBentryALTinterwordspacing
V.~Mnih, K.~Kavukcuoglu, D.~Silver, A.~Graves, I.~Antonoglou, D.~Wierstra, and
  M.~Riedmiller, ``Playing {Atari} with deep reinforcement learning,'' 2013.
  [Online]. Available: \url{https://arxiv.org/abs/1312.5602}
\BIBentrySTDinterwordspacing

\bibitem{rabault2019}
J.~Rabault, M.~Kuchta, A.~Jensen, U.~Réglade, and N.~Cerardi, ``Artificial
  neural networks trained through deep reinforcement learning discover control
  strategies for active flow control,'' \emph{Journal of Fluid Mechanics}, vol.
  865, p. 281–302, 2019.

\bibitem{ren2021}
\BIBentryALTinterwordspacing
F.~Ren, J.~Rabault, and H.~Tang, ``{Applying deep reinforcement learning to
  active flow control in weakly turbulent conditions},'' \emph{Physics of
  Fluids}, vol.~33, no.~3, p. 037121, 03 2021. [Online]. Available:
  \url{https://doi.org/10.1063/5.0037371}
\BIBentrySTDinterwordspacing

\bibitem{fan2020}
\BIBentryALTinterwordspacing
D.~Fan, L.~Yang, Z.~Wang, M.~S. Triantafyllou, and G.~E. Karniadakis,
  ``Reinforcement learning for bluff body active flow control in experiments
  and simulations,'' \emph{Proceedings of the National Academy of Sciences},
  vol. 117, no.~42, pp. 26\,091--26\,098, 2020. [Online]. Available:
  \url{https://www.pnas.org/doi/abs/10.1073/pnas.2004939117}
\BIBentrySTDinterwordspacing

\bibitem{tokarev2020}
\BIBentryALTinterwordspacing
M.~Tokarev, E.~Palkin, and R.~Mullyadzhanov, ``Deep reinforcement learning
  control of cylinder flow using rotary oscillations at low reynolds number,''
  \emph{Energies}, vol.~13, no.~22, 2020. [Online]. Available:
  \url{https://doi.org/10.3390/en13225920}
\BIBentrySTDinterwordspacing

\bibitem{shimomura2020}
\BIBentryALTinterwordspacing
S.~Shimomura, S.~Sekimoto, A.~Oyama, K.~Fujii, and H.~Nishida, ``Closed-loop
  flow separation control using the {Deep Q} network over airfoil,'' \emph{AIAA
  Journal}, vol.~58, no.~10, pp. 4260--4270, 2020. [Online]. Available:
  \url{https://doi.org/10.2514/1.J059447}
\BIBentrySTDinterwordspacing

\bibitem{zheng2021}
\BIBentryALTinterwordspacing
C.~Zheng, T.~Ji, F.~Xie, X.~Zhang, H.~Zheng, and Y.~Zheng, ``{From active
  learning to deep reinforcement learning: Intelligent active flow control in
  suppressing vortex-induced vibration},'' \emph{Physics of Fluids}, vol.~33,
  no.~6, p. 063607, 06 2021. [Online]. Available:
  \url{https://doi.org/10.1063/5.0052524}
\BIBentrySTDinterwordspacing

\bibitem{weiss2016}
\BIBentryALTinterwordspacing
J.~Weiss, Q.~Schwaab, Y.~Boucetta, A.~Giani, C.~Guigue, P.~Combette, and
  B.~Charlot, ``Simulation and testing of a {MEMS} calorimetric shear-stress
  sensor,'' \emph{Sensors and Actuators A: Physical}, vol. 253, pp. 210--217,
  2017. [Online]. Available:
  \url{https://www.sciencedirect.com/science/article/pii/S0924424716309116}
\BIBentrySTDinterwordspacing

\bibitem{weiss2024}
\BIBentryALTinterwordspacing
J.~Weiss and A.~Giani, ``Calorimetric wall-shear-stress microsensors for
  low-speed aerodynamics,'' \emph{Experiments in Fluids}, vol.~65, no.~63,
  2024. [Online]. Available: \url{https://doi.org/10.1007/s00348-024-03803-2}
\BIBentrySTDinterwordspacing

\bibitem{viquerat2022}
\BIBentryALTinterwordspacing
J.~Viquerat, P.~Meliga, A.~Larcher, and E.~Hachem, ``{A review on deep
  reinforcement learning for fluid mechanics: An update},'' \emph{Physics of
  Fluids}, vol.~34, no.~11, p. 111301, 11 2022. [Online]. Available:
  \url{https://doi.org/10.1063/5.0128446}
\BIBentrySTDinterwordspacing

\bibitem{schulman2017}
\BIBentryALTinterwordspacing
J.~Schulman, F.~Wolski, P.~Dhariwal, A.~Radford, and O.~Klimov, ``Proximal
  policy optimization algorithms,'' 2017. [Online]. Available:
  \url{https://doi.org/10.48550/arXiv.1707.06347}
\BIBentrySTDinterwordspacing

\bibitem{henanmemeda2018}
\BIBentryALTinterwordspacing
henanmemeda, ``Rl-adventure-2,'' 2018. [Online]. Available:
  \url{https://github.com/henanmemeda/RL-Adventure-2/blob/master/3.ppo.ipynb}
\BIBentrySTDinterwordspacing

\bibitem{tabor2020}
\BIBentryALTinterwordspacing
P.~Tabor, ``Youtube-code-repository,'' 2020. [Online]. Available:
  \url{https://github.com/philtabor/Youtube-Code-Repository/tree/master/ReinforcementLearning/PolicyGradient/PPO/torch}
\BIBentrySTDinterwordspacing

\end{thebibliography}

\end{document}